\documentclass[journal]{IEEEtran}
\usepackage{graphicx}
\usepackage{xcolor}
\usepackage[normalem]{ulem}
\usepackage{hyperref}

\usepackage{amssymb, amsmath}

\usepackage{amsmath, bm}

\ifCLASSINFOpdf
\else
\fi
\usepackage{amsmath}
\usepackage{array}
\usepackage{fixltx2e}

\usepackage{stfloats}
\begin{document}
%
\title{ The Ka-Band High Power Klystron Amplifier Design Program of INFN}
%
%
%

\author{\IEEEauthorblockN{Mostafa Behtouei$^{1}$,
Bruno Spataro$^1$,
Franco Di Paolo$^{2}$
 and Alberto Leggieri$^{2}$
}

\IEEEauthorblockA{$^1$INFN, Laboratori Nazionali di Frascati, P.O. Box 13, I-00044 Frascati, Italy}

\IEEEauthorblockA{$^2$Dipartimento di Ingegneria Elettronica, Universit\`a degli Studi di Roma \textquotedblleft{Tor Vergata}\textquotedblright, Via del Politecnico, 1-00133-Roma, Italia}


}

\maketitle


\IEEEtitleabstractindextext{%
\begin{abstract}
In the framework of the \textquotedblleft{Compact Light XLS}\textquotedblright project, a short ultra-high gradient linearizer working on the third harmonic of the main LINAC frequency is requested. Increasing gradients and reducing dimensions are requirements for XLS and all next generation linear accelerators. Actually, ultra- compact normal conducting accelerating structures, operating in the Ka-band regime ranging from 100 to 150 MV/m are required to achieve ultra-high gradients for research, industrial and medical applications. To fulfill these strong requirements, the R$\&$D of a proper Ka-band klystron with RF power output and a high efficiency is mandatory. This contribution reports the design of a possible klystron amplifier tube that provides and output signal at 36 GHz to feed the phase space linearizer, while receiving at the input, a low level signal oscillating at 12 GHz; which is the LINAC frequency. The proposed structure reaches an efficiency of 42$\%$ and a 20 MW RF power output. This contribution discusses also the high-power DC gun, the beam focusing channel and the RF beam dynamics.

\end{abstract}

}

\maketitle

\IEEEdisplaynontitleabstractindextext

%
\IEEEpeerreviewmaketitle

\section{Introduction}
%
%
%
%
\IEEEPARstart{H}{igh}-brightness electron beams are required for a great number of applications, including advanced accelerators linear colliders, X-ray Free-Electron Lasers (FELs) and inverse Compton scattering accelerators for research, compact or portable devices for radiotherapy, mobile cargo inspections and security, biology, energy, and environmental applications. This new generation of linear accelerators is highly demanding in terms of accelerating gradients.  In the framework of the INFN-LNF, SLAC (USA), KEK (Japan), UCLA (Los Angeles) collaboration, the Laboratori Nazionali di Frascati (LNF) is involved in the modeling, development and test of short RF structures devoted to the acceleration with high gradient electric field by means of metal devices made with hard copper or copper/silver (Cu/Ag) alloys. A significant fraction of the Research and Development (R$\&$D) has been devoted to studies of new manufacturing techniques to improve the maximum sustainable gradients by short normal conducting hard RF structures. This aim was achieved by minimizing the breakdown and the dark current for extremely demanding applications and for the construction of millimeters high power high frequency cavities \cite{ref1,ref2,ref3,ref4}. In addition, the present R$\&$D using cryogenic copper technology (working conditions $\sim$77 K) will enable a variety of new applications, including linear colliders and Free Electron Laser acceleration, thanks to accelerating gradients over twice the value achieved with the existing technologies \cite{ref5,ref6}.

In the framework of the \textquotedblleft{Compact Light XLS}\textquotedblright \ project \cite{ref7}, a short ultra-high gradient linearizer working on the third harmonic (36 GHz) with respect to the main linac frequency (11.988 GHz) \cite{ref22,ref23,ref24}. In order to operate in the MW power range, an accelerating gradient ~150 MV/m (i.e., 15-20  MV integrated voltage range) is requested \cite{ref25}. To meet these requirements, a 36 GHz pulsed Ka-band RF power source with a pulse length of 100 ns and a repetition frequency  in the (1-10) Hz range is necessary. These frequencies can be easily be reached by solid state devices \cite{ref27,ref28} and innovative techniques \cite{ref29,ref30} are proposed to increase the RF power levels however, actual solution of this kind  are not able to reach MW range of RF power, and vacuum electron devices, especially klystrons are the only solution \cite{ref8}.
It is well known that the klystron amplifier design requires a proper choice of 
\begin{itemize}
\item perveance 
\item beam and pipe diameters  
\item focusing magnetic field 
\item bunching cavities and output cavity system 
\item ultra-vacuum system 
\item coupling coefficient 
\item plasma frequency reduction factor, and  
\item beam collector. 
\end{itemize}
In addition, high electron beam power densities, RF electric  gradient in cavity gaps, and stresses on the ceramic window are just few of the most critical issues to deal with. As a consequence, the choice of the global power source parameters has to be consistent with the obtainable amplifier efficiency and its operation stability.
One of the more challenging aspects of the high power klystron design is the electron gun device. The electron beam perveance is defined as $K = I V^{-3/2}$ where I and V are beam current and voltage. A high perveance results in a lower electronic efficiency due to the higher space charge that affects the beam quality: Since the efficiency is defined as the ratio of the output power to the input one, the voltage is proportional to the $(P_{out}/(\eta K))^{2/5}$. Accordingly, we have to find the best perveance to achieve a satisfactory efficiency with a maneageable voltage \cite{ref8}. 

In this paper, a particular high frequency high power source source is described. In the proposed structure, the bunching section design approach is almost conventional, while the output cavity is an extended interaction cavity \cite{ref8,ref99} allowing loss reduction, breakdowns, etc., while achieving the needed pulsed power \cite{ref9,ref10,ref11}. Consequently, the objective is to design an output circuit, which has the lowest gap fields, consistent with the satisfactory efficiency and the operational reliability. 

A preliminary  theoretical efficiency estimation of the third harmonic klystron operating on the $TM_{01}$ mode is around 18$\%$, about a factor 3 less than the standard klystron efficiency. A design on the 34 GHz Ka-band klystron tube amplifier by achieving more than 10 MW output power with a theoretical efficiency estimation exceeding 22$\%$ has been proposed \cite{ref12}.  Basically, in this design the RF system consists of a X-band input cavity, two X-band bunching cavities, a complicated output cavity and a 5.5 mm diameter beam pipe.  In the output coupler, the output cavity is a $TM_{030}$ mode cavity while others are $TM_{010}$ mode chocke cavities.

The strong motivation of this work is to propose a high power source with a Ka-band harmonic klystron with the output cavity consisting in an extended interaction cavity where 36 GHz is oscillating. This strategy allows to optimize klystron conversion efficiency and provide a peak output power of 20 MW in a 100 ns pulse width, to be used  to feed the linearizer while working  at 150 MV/m of accelerating gradient as requested by all challenging next generation projects \cite{ref6,ref7,ref13,ref14}. For this reason we are also planning to finalize the traveling wave (TW) or the Standing wave (SW) linearizer design as well as the RF power source that will be able to produce up to (50-60) MW input power by using a SLED  system \cite{ref15,ref16}.  

Currently there is no high power klystron operating on 3rd harmonics. It is known that the theoretical efficiency of this device is estimated to be about 1/3 with respect to the standard klystron. The goal of this program is to maximize the klystron efficiency in order to get enough power to feed a short linearizer to be used in Compact Light XlS project  \cite{ref7} and Ultra Compact Cryogenic FEL  \cite{ref5,ref6}. In these applications the linearizers has to be designed in order to provide an integrated voltage at least of 15 MV. To get this integrated voltage we need a power in the range 50-60 MW. By using the SLED system \cite{ref15,ref16} with a compression factor of 4 we can feed the linearizer. The novelty of our work is to design a high power klystron operating at 36 GHz operating to provide enough power to feed the linearizer. It should be noted that with a Gyroklystron, that is able to provide up to 3 MW, also using the SLED system we can achieve 12 MW which is not enough for the projects (Compact Light  lS project and Ultra Compact Cryogenic XFEL).

The paper is organized as follows: section 1 reports the preliminary analysis of the design, section 2 discusses the electron gun and the focusing magnet system, section 3 the interaction structure, the section 4 gives the conclusions of the paper.

\section{Preliminary analysis on the design approach}

Klystrons rely on passing an electron beam through one or more weakly coupled resonant cavities which behaves like grids that serves the purpose of bunching the electrons with resulting high frequency oscillation and amplification of the grid current  \cite{ref12}. The electron beam first interacts with the Buncher cavity, where it undergoes the force related to a low energy alternate field that modulates the electron velocity. As the beam has crossed an opportune distance from the Buncher, the velocity modulation becomes a modulation of the charge density. This charge modulated beam is forwarded into another cavity (the Catcher cavity), where it induces an oscillating field stronger than the field inside the Buncher cavity: from the Catcher cavity the amplified RF field is extracted.

The motivation of this work is to propose a high power source Ka-band klystron in the millimeter wavelength domain for providing a power of 10-20 MW. In designing the klystron amplifier, the choice of a proper electron gun has to be considered, taking with particular attention to achieve a high compression factor with a minimum transverse size. The electron gun relies on the Pierce-type cathode for generating a converging and good laminar electron beam, to be matched with a focusing magnetic field and manipulated in the interaction region by klystron cavities to maximize the RF device power output. Assuming a perveance of 0.3 $10^{-6} A/V^{3/2}$ , a factor of  2 less than other design \cite{ref12},  with  a 480kV pulsed beam voltage, an efficiency of about 42$\%$ can be achieved. 

Since this high voltage could generate breakdowns, an appropriate design in the electron gun regions to withstand these high voltages is mandatory. Cathodes used in standard high power klystrons working in ultra-high vacuum conditions are good candidates. For a long lifetime the cathode loading has to be limited well below 15 $A/cm^2$, too. A cathode diameter in the 8-10 cm range can be considered to this aim. The interaction between the electron beam and the RF cavities system and the Brillouin limit \cite{ref13} are two important parameters in order to control the beam radius limit for design requirements. For getting the optimum klystron efficiency, the beam pipe dimension has to be comparable to the beam radius \cite{ref8}. Moreover, in order to minimize the beam potential depression, a stable and reliable electron beam with a minimal scalloping behavior is also required. 

An average accelerating gradient of 150 MV/m needs at least 50 MW peak RF power. With a 48 MW beam power (480 kV-100 A) and a 100 ns pulse width, 42 $\%$ of klystron efficiency to achieve 20 MW peak RF power on the third harmonic is necessary. Finally, using a SLED system with a compression factor 4 \cite{ref15,ref16}, the RF peak output on the third harmonic for feeding the linearizer is 80 MW.

\section{Electron gun and focusing magnet system design}

The electron gun design has been carried out by using the numerical code CST \cite{ref17}. Intense investigations carried out as function of the geometry anode-cathode, beam current and applied voltage showed a high electrostatic beam compression but with an excessive electric field on the focusing electrodes. As an example, with a 480 kV-210 A beam, or a $\mu$-perveance of 0.631 $A/V^{3/2}$, a 24 MV/m electric field on the focusing electrode, is obtained. Furthermore, in order to obtain a good klystron efficiency, it is also very difficult to make the beam parameters of the klystron to comply to the design requirements. A dedicated and detailed report on the electron gun design will b discussed in a forthcoming paper. However, with a 480 kV beam voltage, 100 A beam current (or a µ perveance of 0.3 $AV ^{-3/2}$), an electrostatic beam compression ratio of 1488:1, the maximum electric field on the focusing electrodes is about 200 kV/cm, a reasonable value for assuring a safety operational margin working with a 100 ns pulse length. Analytical verifications confirmed the electron gun numerical investigation \cite{ref18}. 

\begin{figure*}[t]
 \begin{center}
   \fbox{  \includegraphics[width= 0.98 \textwidth ]{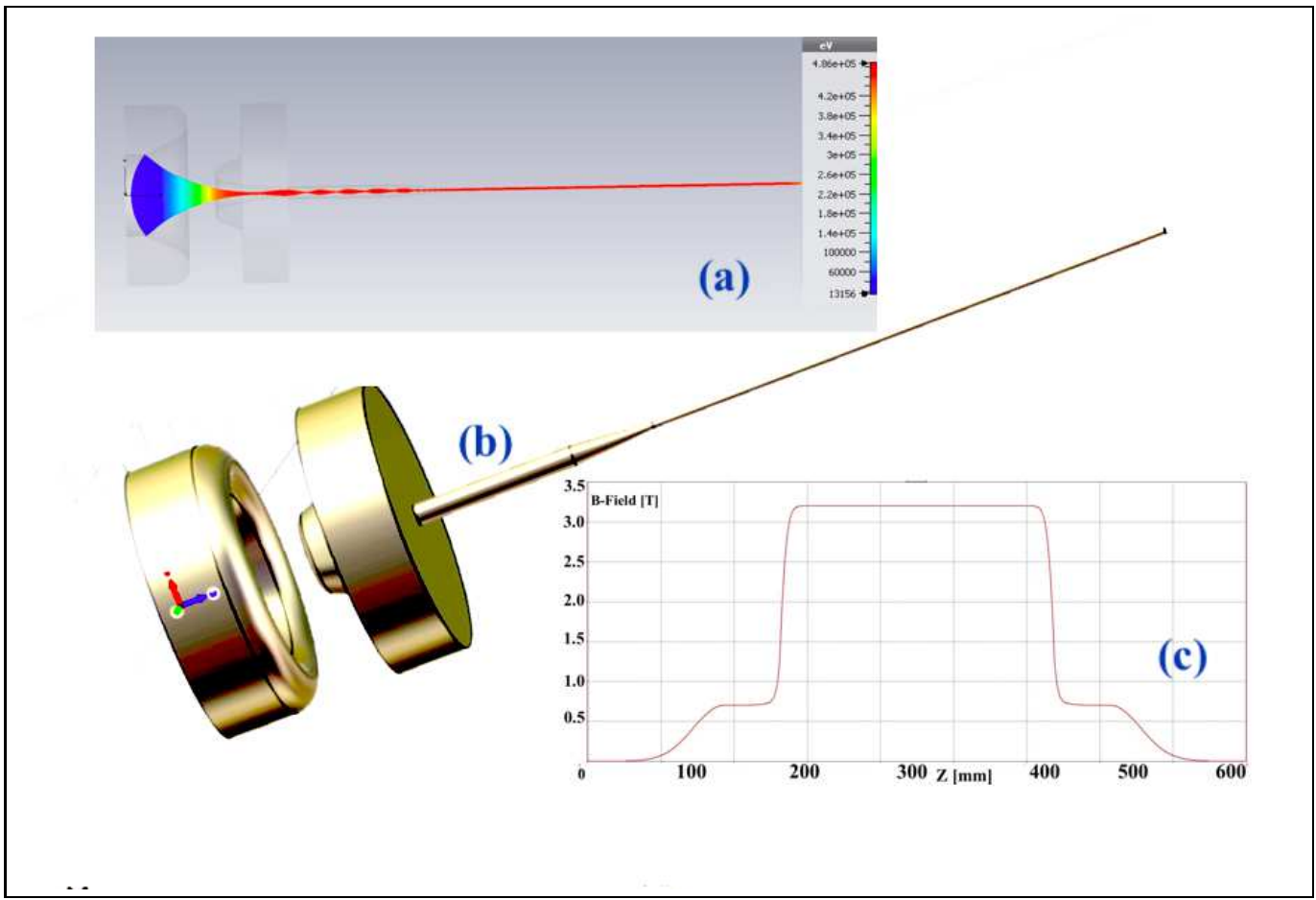}}
      
\caption { a) 3D model of the gun and beam pipe.  b) Beam trajectory along the propagation direction c)  axial magnetic field distribution. }

\end{center}
\end{figure*}

Special attention was given also to the matching among the current generated by the electron gun, the focusing magnetic system and the beam radius dimensions to optimize the beam current transport. The external magnetic field provides both beam focusing and coupling between the electrons and the RF cavities system. Since the output cavity works on the fundamental $TM_{010}$ mode at 36 GHz and its radius is about 3.6 mm, the beam pipe has to be much smaller and comparable to the allowed beam radius, giving klystron efficiency optimization \cite{ref8}.

At the same time, the beam pipe shall place under cut off undesired signals: The drive and the output frequency shall be placed under cut off in the beam pipe. 

Beam tracking simulations of the 480 kV-100 A gun showed that the electron beam is confined within 1 mm radius with a maximum magnetic field of about 32 kG, a 1635:1 beam compression area and 2$\%$ scalloping parameter. The beam pipe reduces the longitudinal space charge, which helps the efficiency in general, but other issues have to be considered as the Beam stratification. We have decided to taper the beam pipe up to 1.2 mm radius to confine the beam radius within 1 mm along the beam pipe to satisfy the beam stratification.

The design parameters of the electron gun, including the magnetostatic beam compression area, are listed in Table 1 and in Fig.1 the 3D model of electron gun, beam trajectory along the propagation direction and magnetic field distribution are reported in Fig.1.
 
\begin{table}[h]

\caption{Design parameters of the gun with focusing magnetic field along the beam axis}

 \begin{center}
\begin{tabular}{||c|c|c||} 
\hline
Design parameters&  \\ 
\hline\hline
Beam power [MW]&  48\\ 
\hline
Beam voltage [kV]& 480\\  
\hline
Beam current [A]  & 100\\ 
\hline
 $\mu-$ perveance $[I/V^{3/2}]$&0.3\\  
 \hline
 Cathode diameter [mm]&  76 \\ 
 \hline
  Pulse duration [$\mu$ sec]& 0.1 \\ 
  \hline
Minimum beam radius in magnetic system [mm]& 0.98\\ 
\hline
Nominal radius [mm]& 1.00\\ 
\hline
Beam pipe radius [mm]& 1.20\\ 
\hline
 Max EF on focusing electrode [kV/cm]&  200 \\ 
 \hline
 Electrostatic compression ratio& 1488:1\\ 
 \hline
Beam compression ratio& 1635:1\\ 
\hline
 Emission cathode current density [$A/cm^2$]&2.02\\ 
 \hline
Beam Transverse Emittance [mrad-cm]& 1.23 $\pi$\\ 
\hline
\end{tabular}
\end{center}
\end{table}

The magnetic field distribution reported in Fig. (1c) shows a small peak of 7 kG and a constant magnetic field of 32 kG along the 300 mm long pipe.This choice allow to achieve  a smaller beam radius suitable to insert the whole system of RF cavities system as it will be described in the next section. In the region where the magnetic field is 7 kG the beam radius is $\sim$2.3 mm, higher by a factor of 4 than the Brillouin limit. In the region where the field is 32 kG, the beam radius is $\sim$ 1 mm, again larger by a factor about 8 than the Brillouin limit, which is $\sim$ 0.13 mm.

We point out also that with the $\mu$-perveance of 0.657 $A/V^{3/2}$, a common value for modern klystron, we could obtain 235 A, although with an efficiency much smaller than with a beam current of $\sim$100 A. Reducing the $\mu$-perveance from 0.657 $A/V^{3/2}$ to 0.3 $A/V^{3/2}$, we maintained the cathode-anode shapes \cite{ref19} while increasing the distance between them in order to obtain a beam current of 100 A maintaining a satisfactory value of the electric field on the focusing electrodes \cite{ref20}.

\section{Interaction Structure Design}

The beam spot is injected in the interaction structure with an initial velocity, given by (1) \cite{ref8}, that is $u_0$ = 2.57 $\times10^8 ms^{-1}$, corresponding to a normalized velocity $u_0$/c = 0.857, where $c_0$ is the speed of the light in the vacuum. 

\begin{equation}
u_0=c_0 \sqrt{1-\frac{1}{(1+\frac{e V_0}{m_0 c_0^2})^2}}
\end{equation}
 
The interaction structure manipulates the beam produced from the electron gun above described with a 12 GHz signal and an input power of 800W.  Our proposed structure is composed by 5 cavities, where the last cavity is an extended interaction cavity integrating 4 gaps \cite{ref8,ref99,ref233,ref21}. It differs from a traditional klystron for both arrangement and shape of the cavities and it uses an extended interaction cavity for the output, where the energy is distributed inside a large volume with the consequent advantage on the surface electric fields.

The structure receives a beam of 100 A and 480 kV with a radius of 1 mm in a drift tube of 1.2 mm radius. 

The first interaction element, encountered by the electron beam, is the Buncher cavity, that operates at 12 GHz and it is followed by gain cavities, designed to have the higher R/Q without degrading the coupling. In the Buncher, the inlet electron wavelength, $\lambda_e = u_o/f$ =21.4 mm, specifies the beam propagation factor $\beta_e = 2\pi / \lambda_e$ =293.5 $m^{-1}$, while the RF wavelength is $\lambda = c_o/f $=25.0 mm, and the wave propagation factor $k =2\pi / \lambda = \omega/c_o$ = 251.5$ m^{-1}$.

While the beam crosses the cavity gap, the velocity of the beam is modulated. The ratio of change in beam voltage is regulated by the beam to wave coupling coefficient {\it M} \cite{ref8}, given by (2). The electric field in the Buncher gap, experienced by electron that traverse the cavity, determines the electron energy variation in the gap. The peak $E_{RF\  pk}$ along the gap, mediated by the coupling factor M, gives a shunt voltage $V_{RF} = M \int_0^d E_{RF\  pk} \ dz$.
The coupling coefficient \cite{ref8} can be numerically calculated by integrating the electric field resulting by an Eigen mode simulation. 

\begin{equation}
M(\beta_e)=\frac{V_{RF}}{V_{RF\  pk}}=\frac{\int_{-\infty}^{\infty} E_z e^{-i \beta_e z} dz}{\int_{-\infty}^{\infty} E_z dz}.
\end{equation}

The mediation is performed in the time interval between the instant when the particle is at the center of the cavity and the instant when the electric field is maximum.

 The beam conductance is $G_0 = I_0/V_0$ = 208.33 $\mu$S and, in the Buncher, the beam-loading conductance, which represents the energy transfer from the cavity to the beam, given by (4) \cite{ref8}, is $G_b = 10.28 \mu S$, that represents a compromise between the need sinusoidal bunching and a non-excessive beam loading.

\begin{equation}
G_b=\frac{G_0}{2} M (M-\cos(\frac{\pi d}{\lambda_{RF}\frac{u_0}{c}})
\end{equation}

\begin{figure}[h]
 \begin{center}
 
   \fbox{ \includegraphics[width= 0.4\textwidth ]{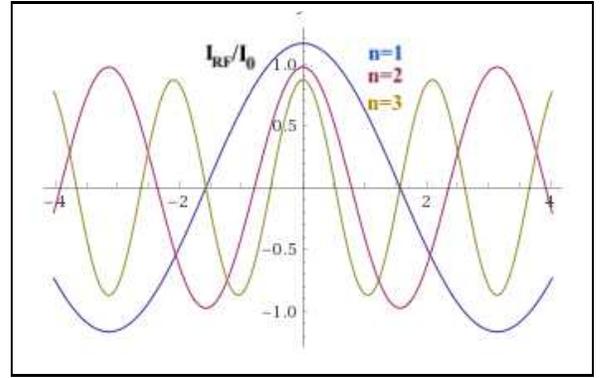} }

\caption { Current modulation depth for the first three harmonics in function of $J_n(X)$ for X = 1.84, 3.05, 4.20 respectively related to the current modulation depth of  $I_1$, $I_2$ and $I_3$.}

 \end{center}

      \end{figure}

The external coupling of the Buncher has been chosen to be reasonably low, in order to use a strong signal at the input, $Q_{ext}$ = 20. This strategy can allow to reduce the energy spread between beam electrons placed at different radial coordinates. The distance between the Buncher and the first gain cavity is chosen to ensure a smooth linear amplification even with a high coupling factor for both the cavities.

The effect on the power transfer from the fundamental to superior harmonics is critically dependent to the bunching parameter X. For a two cavity Klystron it is  \cite{ref8}:

\begin{equation}
X_{Cavity 1\rightarrow 2}=M\frac{V_{RF} (cavity 1)}{2 V_{beam} \frac{2 \pi f \ell }{u_0}}
\end{equation}

The bunching parameter X shall provide, over the full klystron, the needed distortion to enrich the space charge of the 3rd harmonic content, without producing electron reflection or excessive electric fields. The power transfer from the fundamental to the harmonic is function of the Bessel function $J_n(X)$, where n identifies the harmonic index of the considered spectral component of the output signal, as shown in Fig.2. A roughly estimation for the efficiency of the two-cavity klystron operating on the third harmonic is around 19$\%$, about a factor 3 less than the maximum klystron theoretical efficiency operating at the fundamental frequency that is 58$\%$ \cite{ref8}.

      \begin{figure*}[t]
 \begin{center}
  \fbox{   \includegraphics[width= 0.9\textwidth ]{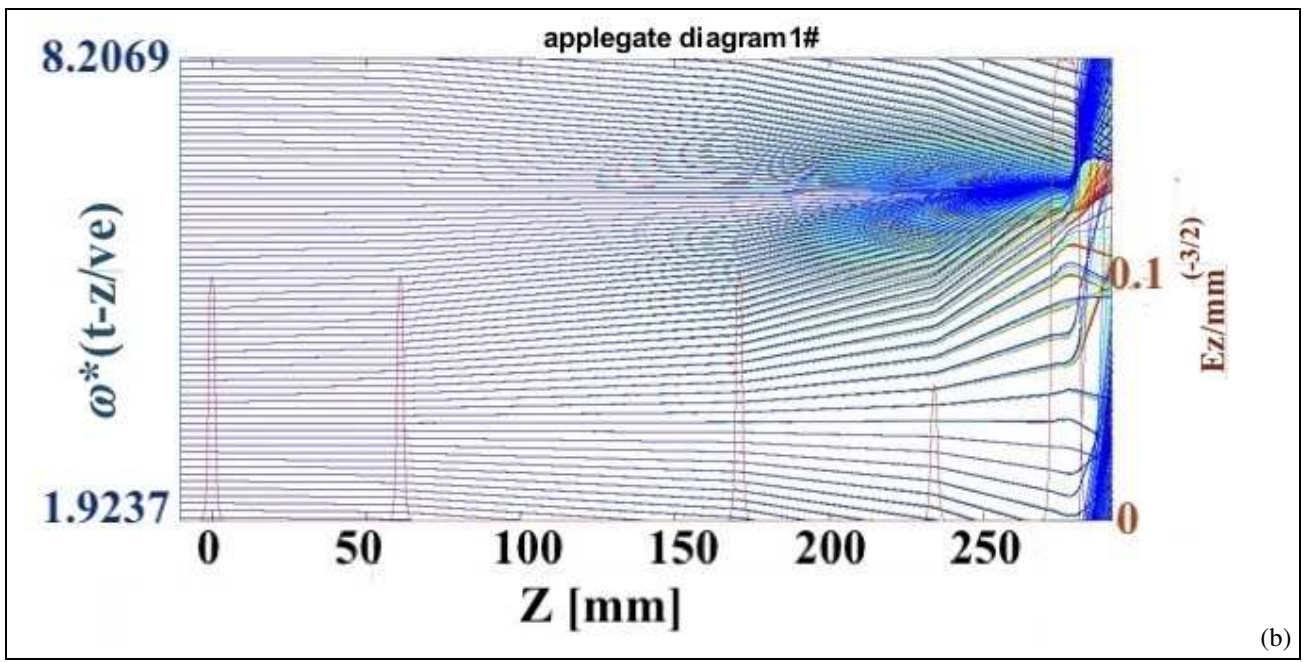} (b)}

\caption {Applegate diagram (distance-time plot) is superimposed to the cavity axial electric field normalized to the maximum value given by the cavity eigenmode calculation.   }

 \end{center}

      \end{figure*}

 \begin{figure*}[h]
 \begin{center}
 
   \fbox{ \includegraphics[width= 0.75\textwidth ]{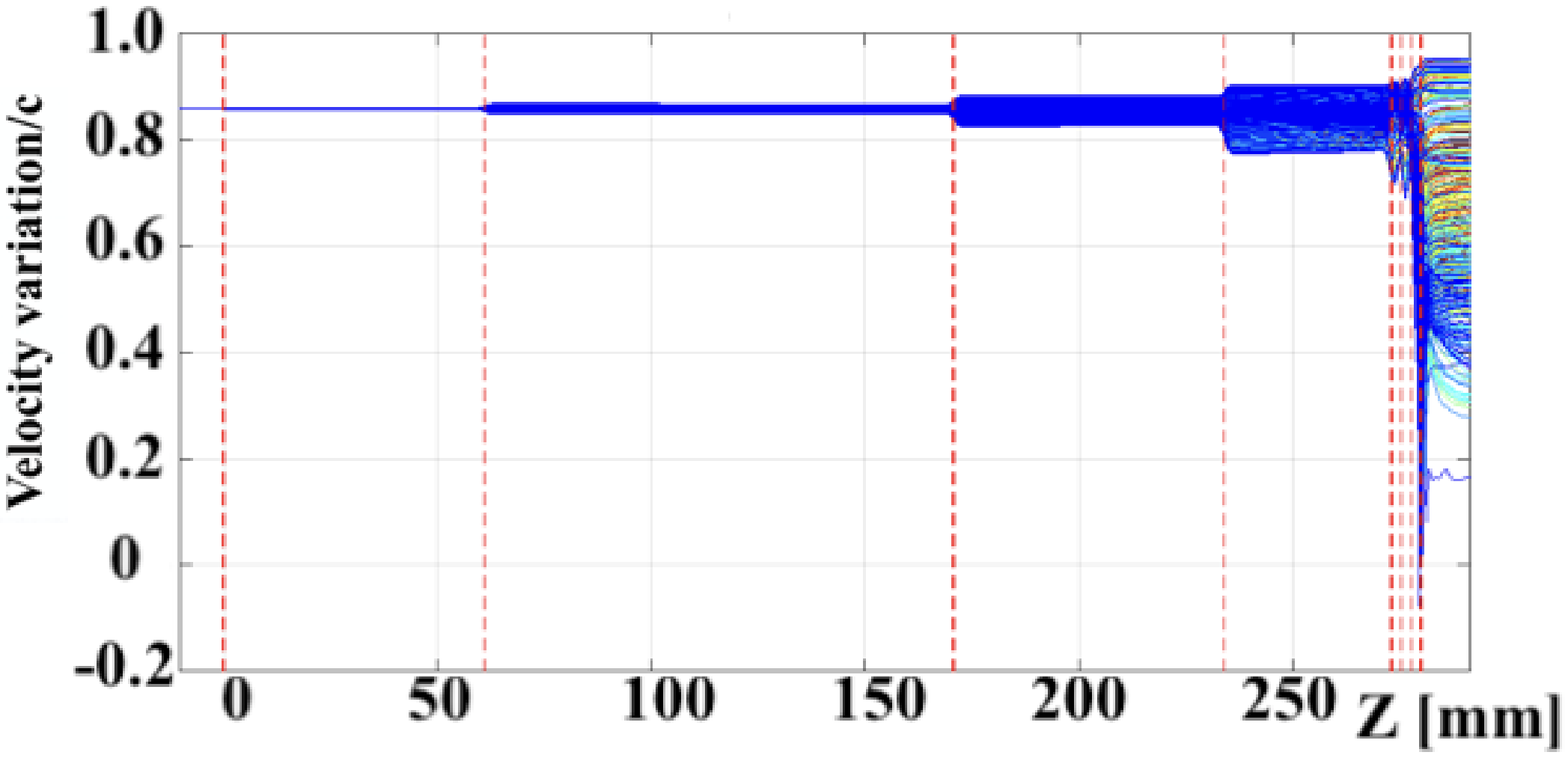} (a)}\\
    \fbox{ \includegraphics[width=0.75 \textwidth ]{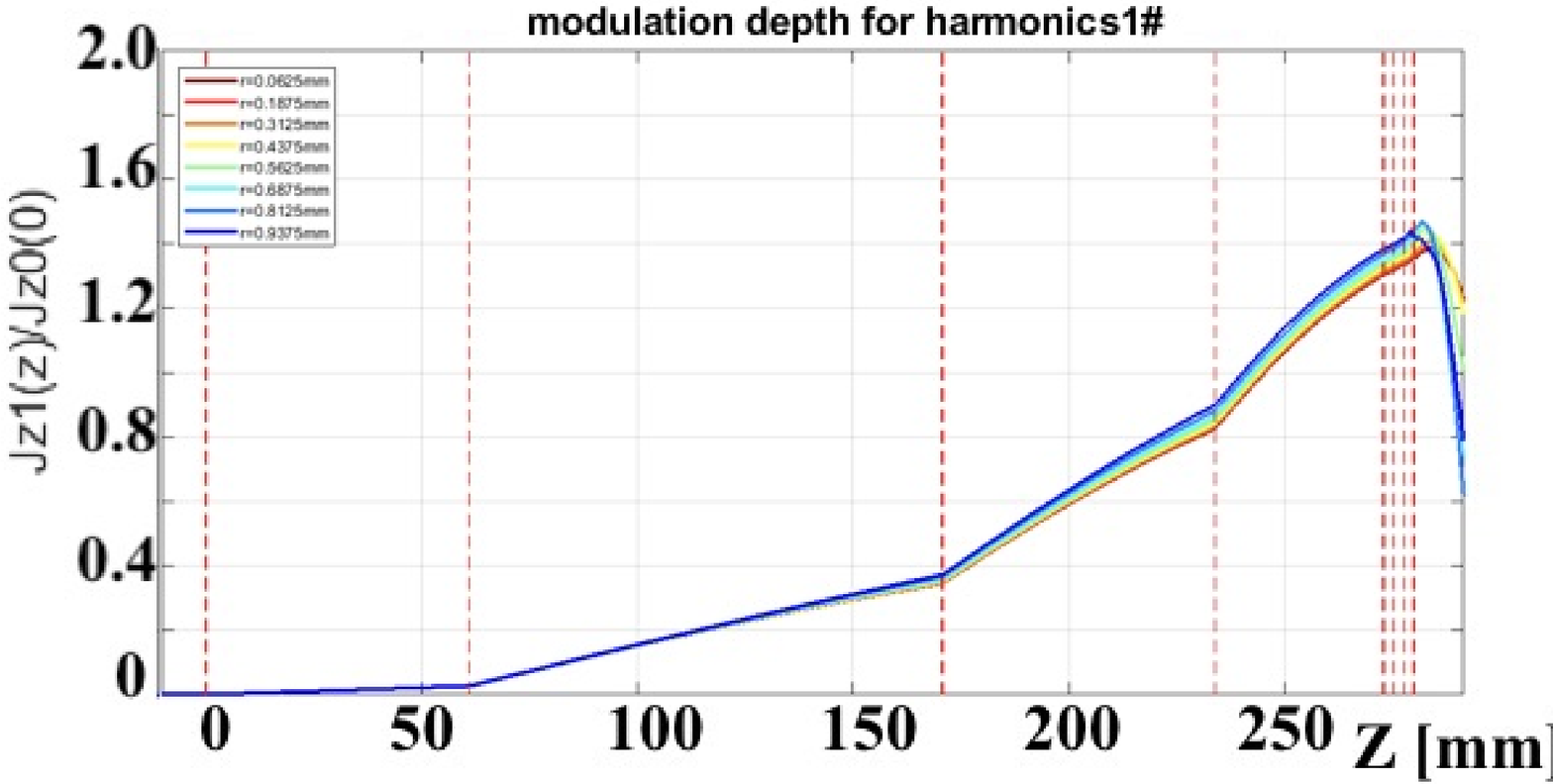} (b)}

\caption { a) The velocity variation;  b) The modulation depth of the 1st harmonic current at different beam substrates as a function of the longitudinal coordinate z. In the vertical axis, $J_{z1}$ is for the 1st harmonic current normalized to the dc current, $J_{z0}$.}

 \end{center}

      \end{figure*}  
 After the Input cavity, the Buncher, three gain cavities are placed with the opportune drift spaces in order to operate a smooth amplification and to produce the desired signal compression before the output cavity. The gap of the gain cavities is enough large to have a good coupling with the beam and ensure the desired signal growth.
The Output circuit is an extended interaction output cavity that integrates four gaps \cite{ref233,ref21}. The multiple gaps are designed to be synchronous with the beam for the mode chosen during the passage of the beam inside the whole cavity. This output circuits do not obey this rule and make possible wider bandwidths than that of a traditional cavity and/or more power extraction from the beam.
The beam, that traverses the extended interaction cavity, is subjected to a continuous extraction of energy during the whole path.  Such a cavity has not only a larger surface from which to remove the heat, but also a higher R/Q, which leads to a lower eternal quality factor and a higher circuit efficiency. Within the extended interaction cavity, the restrictions of single or gridded gaps are removed and the beam interaction takes place with a generalized field. Such a field may consist of several discrete regions, corresponding to cavities placed closely together, or it may be the continuous field of a shorted helix or a coupled cavity structure  \cite{ref8}.

The most important aspect is that these structures for R/Q values that cannot be attained with single-gap cavities. As clearly explained by Caryotakis \cite{ref8}, this is a direct result of the fact that these circuits are no longer driving-point impedances as single-gap cavities are and hence not subject to the bandwidth limitation. Extended Interaction klystrons can attain bandwidths modestly superior to ordinary klystrons.
High R/Q output circuits also imply low external $Q_{ext}$, and higher circuit efficiency. Extended interaction cavities can distribute the necessary retarding RF voltage at an output cavity over several gaps, thereby reducing the RF gradient and the danger of RF breakdown. This is the principal reason for the use of extended cavities in pulsed klystrons, though circuit efficiency was also a consideration \cite{ref8}. In the presented structure, from the Extended Interaction Cavity, the signal at the output frequency is provided. 

The R/Q, coupling coefficient and $G_b/G_0$ for an extended interaction cavity must be chosen carefully and the beam loaded conductance must be positive for the operating mode. This approach required a certain value of micro-perveance to produce a satisfactory efficiency. The gap to gap period is dimensioned to ensure the synchronism between the beam and wave phase velocity, while the beam traverse the cavity. Depending on the beam voltage, the operative cavity mode is selected \cite{ref8}. An inconvenient of the extended interaction cavities is the possibility to lead oscillations at lower frequencies than the operating.

Interesting solutions are being proposed in the domain of extended interaction techniques to be used in frequency conversion mostly by using higher order mode output cavities, as proposed by G. Burt \cite{ref211} and D. Consatble \cite{ref222} at Lancaster University and J. Cai and I. Syratchev\cite{ref233} and CERN. 

An important aspect concerns the dimension of the structure that determines the possibility to be manufactured.

 \begin{figure*}[t]
 \begin{center}
 \fbox{\includegraphics[scale=0.55]{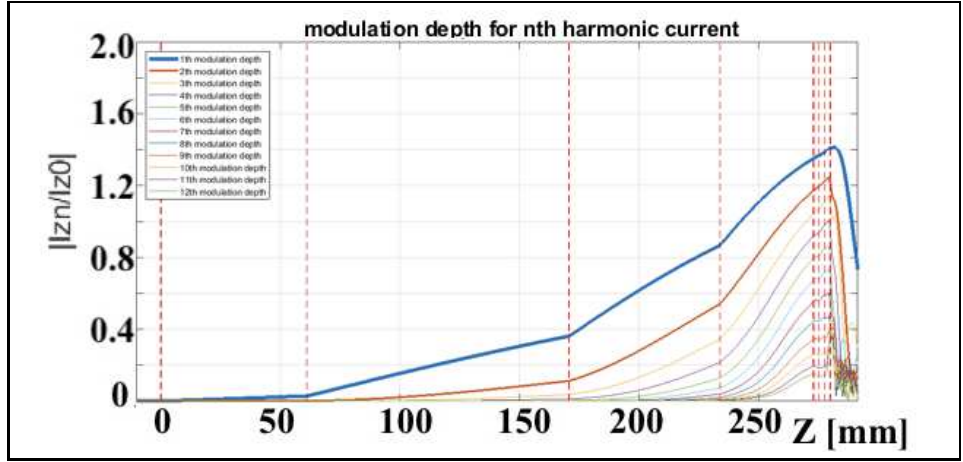}}

\caption{The modulation currents normalized to the dc current, $J_{z0}$ as function of the longitudinal coordinate z. In the vertical axis $J_{zn}$ is the higher harmonic current, which is calculated for different harmonic (n=1,2,..,12). }
\end{center}

  \end{figure*}  
The klystron Buncher is a cylindrical cavity with a radius of 8.43 mm, operating in a quasi $TM_{010}$ mode. The beam pipe has a radius aperture of 1.2 mm. The extended interaction \cite{ref99} output cavity has a radius of 3.37mm can be realized using Electrostatic Discharge Machining (EDM). The size of the component composing the designed structure allows for manufacturing processes based on Control Numeric Computer (CNC) machining.
Klystron performances are limited by the beam voltage reduction effect and by the more invasive large signal phenomena, including plasma resonances; which introduces oscillations below the drive frequency, as given by (6) \cite{ref8} that apply a plasma reduction factor on the interaction frequency.

\begin{equation}
f_p=\frac{1}{2\pi} \sqrt{\frac{e}{m_0} \frac{J_0}{\epsilon_0 u_0 \gamma^3}}
\end{equation}

where $e/m_0=1.759 \times 10^{11}$ is the electron charge to rest mass ratio, $\epsilon_0=8.85\times 10^{-12}$ F/m the permittivity of vacuum and $\gamma$ = 1+ V0 / 511kV is the Lorenz factor and u0 the beam velocity. In presence of deep modulation $V_{RF}>>V_0$, the space charge effect due to the plasma frequency reduces the bunching parameter. A numerical approach for a large signal deign is mandatory.

Numerical behavioral characterization of high efficiency klystron amplifier in large signal regimes requires specialized computer codes. 1D codes like AJDisk \cite{ref2111}, and KlypWin, operate in an idealized environment, allow for a very fast (1-10 min) computation time. 2D codes like TESLA, Klys2D and FCI, account for all radial effects and economize on computation time (10 min - 1 hour). Particle-In-Cell (PIC) codes like MAGIC2D/3D and CST Particle Studio provide complete and self-consistent field solution therefore the computation time could be very large (10-100 hours) \cite{ref21}. A new 1D/2D complex simulation code has been developed by Dr. Cai and Syratchev at CERN to operate with a precision close to that of a PIC code with the benefit of discrete approach that reduces the computation time to 5-30min. The used approach assumes that radial beam expansion has little effect on the RF power production; therefore, only longitudinal beam dynamic is enough to be considered in the klystron optimization routine. In such approximation, PIC approach can be successfully replaced by the model based on the generic large signal theory. Such a theoretical model was developed and implemented into new large signal 1D and 2D klystrons simulation code called KlyC \cite{ref21}. It has been demonstrated by a large number of examples on high efficiency klystron simulations, operating in very difficult computational condition needed to get high efficiency that the KlyC code is equivalent to a PIC code when cavity geometries or cold fields are used as source data for the beam dynamics computation. The KlyC simulations were evaluated by comparison with the CST PIC code, and both codes showed a good (within 1$\%$) agreement, while KlyC was significantly (about 100 times) faster \cite{ref21}.

The software KLyC \cite{ref21}. has been used for the computer aided design of the structure. The best effort has been done to obtain a sufficiently short structure suitable to accommodate a considerably narrow focusing magnet and obtain the perveance needed for the required beam power output ensuring the maximum efficiency.

The presented simulations have been obtained by setting the following KlyC parameters \cite{ref21}.: Number of beams: 8, Space charge field order: 12, Electron wavelength division number: 256, RF wavelength division number: 128, Iteration residual limit:10-4, Max iterations: 64, Iteration relaxation: 0.25.

At the same time, a minimal velocity, below -0.1c, has been considered to avoid electron reflections in the output cavities. The phase grouping is shown in Fig. 3, where the Applegate diagram is superimposed to the cavity axial electric field normalized to the maximum value given by the cavity Eigen-mode calculation. In order to get high efficiency, the electron beam is significantly decelerated in the last cavity of the output coupler down to a velocity of -0.09 c along the z direction. This effect occurs in the last cavity, where the beam, has a considerably reduced energy. The velocity variations are showed in Fig. (4a). The modulation depth of the $1^{st}$ harmonic current at different beam substrates is shown in Fig. (4b). In this figure, $J_{z1}$ is the $1^{st}$ harmonic current normalized to the dc current, $J_{z0}$. The 1st harmonic content has a specific distribution on different substrates of the beam. The inner substrates have a slightly reduced fundamental content and this is one of the reason for which the last gain cavity is a low R/Q cavity.

The superior harmonic content grows while the signal is sufficiently compressed; this operation is demanded mostly to the $4^{th}$  cavity. In the output cavity, the $3^{rd}$ harmonic phasor of the drive frequency, reaches the amplitude 0.77 of the amplitude of the fundamental. The output signal oscillates at 36 GHz. In Fig. (5), $J_{zn}$ is the higher harmonic current, which is calculated for different harmonic (n=1,2,..,12). The output power is 20 MW at 36 GHz. The max gap voltage is 471 kV, which is produced in the extended interaction output cavity, where is also located the maximum electric field (222 MV/m).
 
The minimum velocity level is  $-0.09^{*}c$; which fulfils the requirement on $v_{min}>-0.1^{*}c$. A summary of the klystron output is reported in Table II. The considered efficiency is the Balanced Efficiency computed by KlyC. This quantity accounts for the ohmmic losses that, a traditional discrete method can't.

 \begin{table}[h]

\caption{Resume of Klystron Output}

 \begin{center}
{  \begin{tabular}{|c|c|c||} 
\hline
Simulated quantity& Output data \\ 
\hline
$P_{out}$ [kW]&  19983.804\\ 
\hline
Gain [dB]& 43.976\\  
\hline
RFefficiency  & 0.416\\ 
\hline
 Min v/c&-0.094\\  
 \hline
\end{tabular}}
\end{center}

\end{table}

The design software KLyC has been used for the computer aided design of this structure. With this tool efforts have been made to design a sufficiently short structure to be accommodated inside a considerably narrow focusing magnet, characterized by a high perveance as required by the beam power output, while ensuring the maximum efficiency. At the same time, a minimal velocity, below -0.1 c, has been obtained to avoid electron reflections in the output cavities. The output element, a waveguide iris facing on a dielectric window, can be designed. Since the Klystron produces a higher signal at the X band; the window can be optimized at the Ka band, in order to not reduce power in this range.

Two dielectric windows could be also designed following standard rules \cite{ref26}. The max gap voltage is 302.7 kV and the max electric field is 138.4 MV/m that for a 200ns pulse avoid electric discharges.

\section{Conclusions}

This paper proposes a novel Klystron structure that provides an output signal to energize a phase space linearizer while receiving at the input the LINAC frequency.

In order to linearize the longitudinal space phase of the Compact Light XLS project a Ka-Band accelerating structure operating on the $\pi$ mode at the third RF harmonic with respect to the main LINAC RF frequency has been considered. The linearizer can work with a high accelerating gradient around up to 150 MV/m. In this contribution, a klystron amplifier has been also investigated to feed this linearizer structure. We presented the design of the high-power DC gun, of the beam focusing channel and of the RF beam.

The numerical model has been applied to the needs of the Compact Light XLS project, to design a high power Klystron multiplier optimized to provide a 36 GHz output while receiving a 12GHz input.

The proposed structure can be supplied by same Low Level Radio Frequency driver of an X-band klystron, offering the possibility to control the phase of the two power source outputs. Lower parasitic and phase noise than using a single higher frequency driver would be also present. The proposed structure shows an efficiency of 42$\%$ and a gain of 44dB. At the output of the device a signal at the 3rd harmonic of the drive frequency is generated with a power of 20 MW.

The max gap voltage is 302.7 kV and the max electric field is 138.4 MV/m that for a 200ns pulse avoid electric discharges. The minimum velocity level is -0.094 c, which respect the common design requirement on $v_{min}>$-0.1c, to prevent electron reflections. This device is particularly suitable to produce pulsed signals ranging from 100 to 200 ns.

This feature is in line with requirements of high gradient accelerators operating in Ka band, which are foreseen to achieve gradients around 150 MV/m. 

The dimension of the cavities, will allow using a complete CNC process without the use of electro erosion technology, needed to machine the cavities. 

The application of the proposed design principle is very large: In addition to the 12 GHz to 36 GHZ application, it can be efficiently used in a frequency manipulation from 4 to 12 GHz. In order to extend the applicability to higher frequencies of the proposed principle or to reduce machining complexity, structures using higher order mode cavities can be engineered

\section*{Acknowledgment}
The authors would like to thank Cai Jinchi, Igor Syratchev, and Zening Liu for their support in the development of the RF beam dynamic. 

This work was partially supported by the Compact Light XLS Project, funded by the European Union's Horizon 2020 research and innovation program under grant agreement No. 777431.

%








\end{document}